\documentclass[12pt,a4paper]{article}

\usepackage{amsmath}
\usepackage{graphicx}
\usepackage{times}
\begin{document}
\begin{titlepage}
\title{Reflective scattering mode interpretation}
\author{ S.M. Troshin, N.E. Tyurin\\[1ex]
\small  \it NRC ``Kurchatov Institute''--IHEP\\
\small  \it Protvino, 142281, Russian Federation,\\
\small Sergey.Troshin@ihep.ru\\
\small PACS: 13.85.Dz; 21.60.Ev}
\normalsize
\date{}
\maketitle

\begin{abstract}
This note is devoted to discussion of  the reflective scattering mode. This mode connection  to the color conducting medium formed under hadron collisions is proposed.
\end{abstract}
\end{titlepage}
\setcounter{page}{2}
\section{Introduction}
Hadrons are  extended composite objects, i.e. their formfactors are nontrivial functions. There is a  naturally expected  contribution of inelastic interactions increasing with energy,  but simultaneously a significant contribution to $pp$--interactions is provided by the elastic scattering , the ratio of the elastic to total cross-sections $\sigma_{el}(s)/\sigma_{tot}(s)$ is  rising with energy. It means that the ratio of $\sigma_{inel}(s)/\sigma_{tot}(s)$ decreases with the collision energy growth. 

Importance of the elastic scattering for strong interactions dynamics  was underlined long time ago by Chew and Frautchi \cite{chew} in the study based on the Mandelstam representation.  They noted that a ``characteristic of strong interactions is a capacity to ``saturate'' the unitarity condition at high energies''.  It provides a hint on the dominant contribution of the elastic scattering {\it decoupled from the multiparticle production}   due to unitarity saturation at $s\to\infty$ . 

 It should also be noted here, fortifying what was said  above,  that the upper bound for the inelastic cross--section obtained recently  \cite{wu} {\it excludes}  $\ln^2 s$-dependence for $\sigma_{inel}(s)$ at $s\to\infty$ when the ratio
 $\sigma_{tot}(s)/\sigma_{tot}^{max}(s)$, where $\sigma_{tot}^{max}(s)=(4\pi/{t_0})\ln^2(s/s_0)$,  follows its limiting behavior, i.e. it tends to unity at $s\to\infty$. Here $\sigma_{tot}^{max}(s) $ represents the Froissart--Martin bound for the total cross-sections and results from saturation of unitarity  for the elastic scattering amplitude and its analytic properties.
 
Extrapolation of the observed experimental dependencies to higher energies is performed in the two  ways: one can assume equipartition of the elastic and inelastic contributions at $s\to\infty$ (black disc) or allow  saturation of the unitarity bound leading to the asymptotic behavior of the ratios $\sigma_{el}(s)/\sigma_{tot}(s)\to 1$ and $\sigma_{inel}(s)/\sigma_{tot}(s)\to 0$. Some intermediate dependencies of these ratios at $s\to \infty$ can also be envisaged (cf.  \cite{fagund}). Both of the above mentioned modes are consequences of unitarity, but rejection of the reflective scattering by assuming $f(s,b)\leq 1/2$ (where $b$ is an impact parameter of the colliding hadrons, note that $l=b\sqrt{s}/2$) means {\it ad hoc} limitation imposed on the wealth of the dynamics provided by the unitarity. 

The reflective scattering mode corresponds to the unitarity saturation option. The main feature of this  mode is negativity of the elastic scattering matrix element $S(s,b)$ leading to  the asymptotic dominance of the reflective elastic scattering  and the respective peripheral character  of  the inelastic scattering overlap function in the impact parameter space.
Decoupling of  the elastic scattering  from the multiparticle production occurs initially at small values of the impact parameter $b$ expanding to larger values of $b$ with increase of energy. Such a behavior corresponds to increasing self--dumping of  inelastic contributions to unitarity equation \cite{baker}. The knowledge of the decoupling dynamics  is essential   for  the hadron interaction's studies, e.g. for the  development of QCD in its nonperturbative sector where the color confinement plays a crucial role. 

The $b$--dependence of the scattering amplitude as well as of the inelastic overlap function should be considered as a collision geometry.
It should  be emphasized that the collision geometry describes   the  hadron interaction region  but not  the spacial properties of the individual participating hadrons.
\section{Reflective scattering mode}
We recall now the main features of the reflective scattering mode. Partial wave matrix element of the elastic scattering  is related to the corresponding amplitude $f_l(s)$ by the relation 
\[
S_l(s)=1+2if_l(s),
\]
where the amplitude $f_l(s)$ obey the unitarity equation:
 \begin{equation}
 \mbox{Im} f_l(s)=|f_l(s)|^2+h_{l,{inel}}(s). \label{ub}
 \end{equation}
It is convenient to use an impact parameter representation, which 
 provides a simple semiclassical picture of hadron scattering, recall that $l=b\sqrt{s}/2$.
For simplicity, we use a common assumption on the smallness of the
 real part of the elastic scattering amplitude  in the impact parameter representation $f(s,b)$  and perform replacement $f\to if$. It should be noted that this assumption correlates with unitarity saturation. Indeed,  unitarity saturation means that $\mbox{Im} f(s,b)\to 1$ at $s\to\infty$ and fixed $b$. It can easily be seen that this limiting behavior implies that  $\mbox{Re} f(s,b)\to 0$ at $s\to\infty$ and fixed $b$ and leads to inconsistency \cite{pl} of Maximal Odderon \cite{nic} with unitarity saturation.

Unitarity equation provides the evident relation for the dimensionless differential distribution over $b$ of the inelastic collisions $h_{inel}(s,b)$  in  case of proton--proton scattering
\begin{equation}\label{pinel}
h_{inel}(s,b)= f(s,b)(1-f(s,b)).
\end{equation}
It  constraints variation of the  amplitude $f(s,b)$ by the values from the interval
$0 \leq f \leq 1$. The value of $f=1/2$ corresponds to the complete absorption of the initial state and means that the elastic scattering matrix element  is zero, $S=0$  (note that $S=1-2f$).  If the amplitude $f(s,b)$ at $b=0$ (beyond some threshold value of energy) becomes greater than $1/2$,  then the maximal value of differential distribution of inelastic collisions is $1/4$ at $b> 0$ (cf. Eq. (\ref{pinel})).  Thus, approaching unitarity saturation limit in the region where $f>1/2$ leads to a peripheral nature of the inelastic hadron collisions' differential distribution over impact parameter. Such peripheral character is a straightforward result of a probability conservation, i.e. unitarity.

 Reflective scattering mode corresponds to the region of the amplitude $f$ variation in the range $1/2 < f \leq 1$. It means that $S$ is negative and varies in the region $-1\geq S < 0$.
 The negativity of the elastic scattering matrix element is the reason for the term "reflective scattering". Its interpretation will be discussed in the next section.

The value of the collision energy corresponding to the complete absorption of the initial state
under the central collisions  $S(s,b)|_{b=0}=0$
is denoted as $s_r$ and the estimates for the value of $s_r$ are of order of  few  $TeV$ \cite{srvalue}. This is confirmed by the impact parameter analysis performed in \cite{alkin}.
At the energies $s\leq s_r$ the scattering in the whole range of impact parameter variation
has a shadow nature (Fig. 1), it means that solution of the unitarity equation for the  elastic  amplitude has the form:
\begin{equation}\label{shad}
f(s,b)=\frac{1}{2}[1-\sqrt{1-4h_{inel}(s,b)}], 
\end{equation} 
which assumes a direct coupling of elastic scattering to multiparticle production often called shadow scattering.
When the  energy value becomes larger than 
 $s_r$, the scattering picture at small values of impact parameter 
($b\leq r(s)$, where  $S(s,b=r(s))=0$) starts to acquire a reflective contribution. At such energy    and impact parameter values  unitarity gives for the  elastic  amplitude  another form:
\begin{equation}\label{ashad}
f(s,b)=\frac{1}{2}[1+\sqrt{1-4h_{inel}(s,b)}].
\end{equation}
\begin{figure}[hbt]
	\vspace{-0.42cm}
	\hspace{-2cm}
		\resizebox{16cm}{!}{\includegraphics{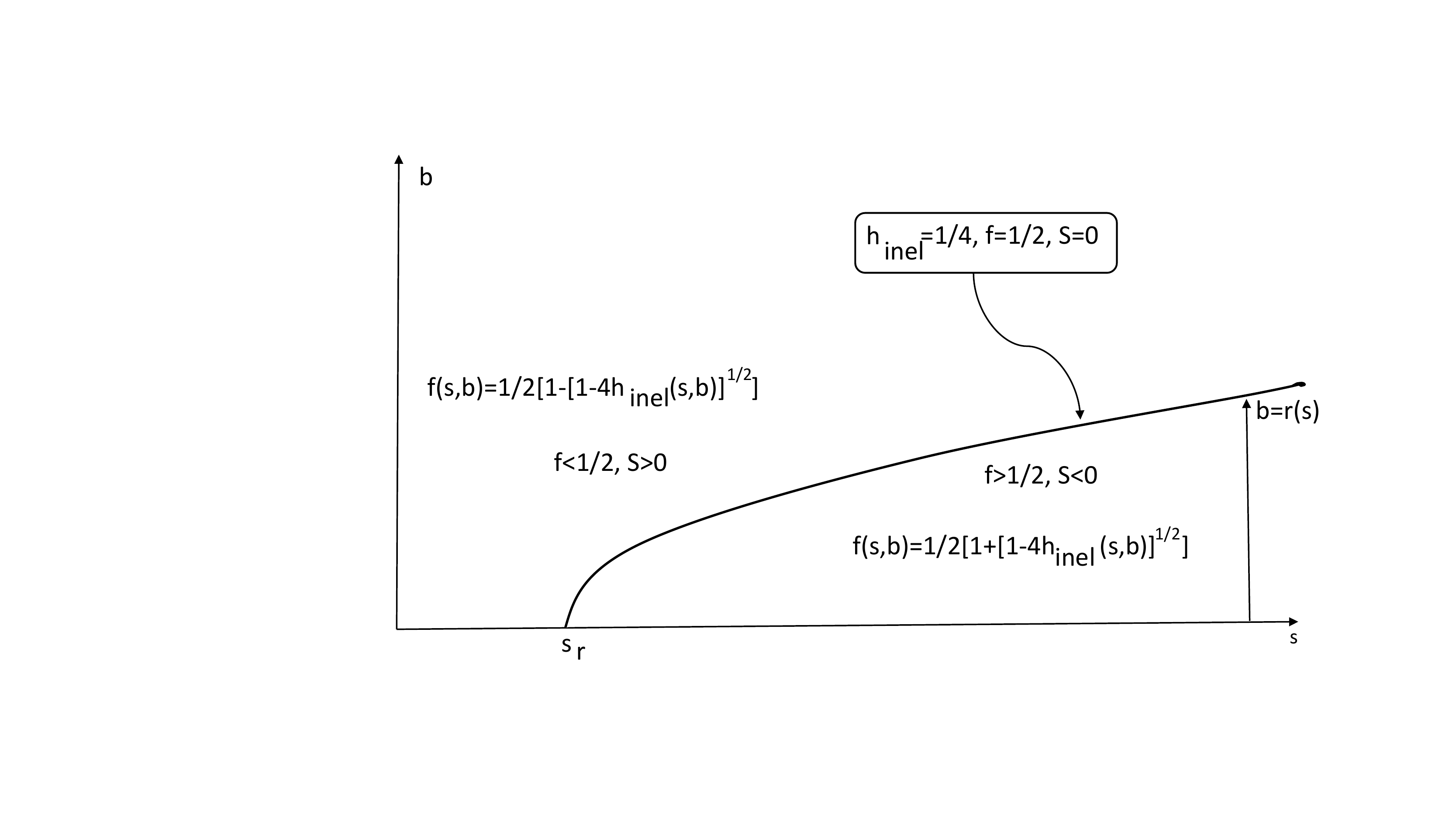}}		
	\vspace{-2cm}
	\caption{Schematic representation of the regions in $s$ and $b$ plane corresponding to absorptive ($S>0$) and reflective ($S<0$) scattering modes.}	
\end{figure}	 
Eq. (\ref{ashad}) corresponds to growing decoupling of the elastic scattering from multiparticle production. It can exist in the limited range of  the impact parameter values only, namely, at $b\leq r(s)$ since at larger values of $b$: $f\sim h_{inel}$. 
At $s>s_r$ the function $h_{inel}(s,b)$ has a peripheral dependence on $b$\footnote{It is claimed \cite{samo} that account of the real part of the elastic scattering amplitude $f$ leads to the central dependence of $h_{inel}$ on $b$. Despite  such possibility could be possible in principle (since $h_{inel}$ transforms into $h_{inel}+(\mbox{Re} f)^2$ in the unitarity equation) this option is not realized in practice. Impact parameter analysis performed in \cite{alkin} {\it with account of the real part} leads to the peripheral form of $h_{inel}$ dependence on $b$. Therefore, account for the real part does not alter peripheral form  of $h_{inel}$ and  neglect of its contribution is  justified at least qualitatively.} since
\begin{equation}
\label{der}
\frac{\partial h_{inel}(s,b)}{\partial b}=S(s,b)\frac{\partial f(s,b)}{\partial b},
\end{equation}
where $S(s,b)$ is negative at $s>s_r$ and $b< r(s)$.

\section{Reflective scattering mode interpretation}
The  elastic scattering matrix element can be written in the form
\begin{equation}
S(s,b)=\kappa (s,b)\exp[2i\delta(s,b)].
\end{equation}
Here $\kappa $ and $\delta $ are the real functions and $\kappa$ can vary in the interval
$0\leq \kappa \leq 1$.  This function is called an absorption factor and its value $\kappa =0$ corresponds to complete absorption of the initial state, 
\begin{equation}
\kappa^2 (s,b)=1-4h_{inel}(s,b).
\end{equation}
The function $S(s,b)$ can be nonnegative in the whole region of the impact parameter variation or have negative values in the region $b<r(s)$ when the energy is high enough, i.e. at $s>s_r$. 
Under the reflective scattering, $f>1/2$, an increase of elastic scattering amplitude $f$ corresponds to decrease of $h_{inel}$ according to 
\[
(f-{1}/{2})^2={1}/{4}-h_{inel}
\]
and, therefore, the term antishadowing has initially been used for description of such scattering mode emphasizing that the reflective scattering is correlated with the self-damping of the inelastic channels contributions \cite{baker} and increasing decoupling of the elastic scattering from multiparticle production dynamics.

Transition to the negative values of $S(s,b)$ means that the  phase $\delta$ changes its value from $0$ to $\pi/2$.  The term reflective is borrowed from optics
since the phases of incoming and outgoing  waves differ by $\pi$. It happens when the reflecting medium is optically denser (i.e. it has  a higher refractive index than the  medium where incoming wave travels before impinging the scatterer). 
Thus, there is an analogy with the sign change  under reflection of the electromagnetic wave by a surface of a  conductor. This occurs because of  the  electromagnetic field  generates a current in this medium. 

 To consider  the reflective scattering mode interpretation it is convenient to use a ``quantum mechanical laboratory''  (cf. \cite{schr}).  A standard  separation of the c.m. motion allows one to reduce the case to the stationary problem of a plane wave scattering   off the  scatterer, representing the interaction region. Evidently, the plane wave and the scatterer should not be identified with the incoming hadrons. 
The energy evolution of the scatterer   leads to appearance of  the reflective scattering mode if one admits the unitarity saturation in the limit of $s\to\infty$. 

Reflective scattering mode does not assume
any kind of transparency during  the head-on collisions. Contrary, it is about the geometrical elasticity  (cf.\cite{usprd}). The interpretation of the reflective scattering mode based  solely on the consideration of inelastic overlap function  is, therefore, a deficient one. It can lead to incorrect interpretation based on idea of the formation of the hollow fireball (e.g. filled by the disoriented chiral condensate) in the intermediate state of hadron--hadron interaction (cf. \cite{blr}) and considering  the central region as  the transparent one. It should be noted that term transparency itself  can only be relevant for the energy and impact parameter region responsible for the shadow scattering regime, i.e. where $f<1/2$ (cf. Fig 1).

The emerging physical picture of   high energy  interaction region  in transverse plane can be interpreted then
as  a reflecting
disk (with albedo approaching to complete reflection at the center at $s\to\infty$) which is surrounded by a   black ring 
(with complete absorption, $h_{inel}=1/4$) since the inelastic overlap function $h_{inel}$ has a prominent peripheral form in this scattering mode.
The reflection  mode implies that the following  limiting behavior $S(s,b)|_{b=0}\to -1$ will take place at $s\to\infty$\footnote{Despite the limiting behavior of $S(s,b)$ corresponds to $S\to -1$ at $s\to\infty$ and fixed $b$, the gap survival probability, contrary to conclusion of  \cite{khoze}, tends to zero at $s\to\infty$ (cf. \cite{gsp}).} . Of course, it is supposed a monotonic increase  of the amplitude $f$ with energy to its unitarity limit unity, and an artificial option  of its nonmonotonic energy dependence at fixed values of $b$ \footnote{Such nonmonotonic behaviour might result (after integration over $b$)  in peculiar distortions superimposed onto the rising  energy dependence of the total cross-section.} is excluded.

QCD is a  theory of hadron interactions  with colored objects as the sources and colored  mediators of the interaction. Thus, one can imagine that when the energy of the interacting hadrons increases beyond some threshold value the color conducting medium being formed as a result of a color deconfinement. Properties of such medium are under active studies in nuclear collisions, but color deconfinement,  can take place in hadron interactions, too. Therefore, it is  tempting  to relate  appearance of the reflective scattering mode   with formation of the color conducting medium in the intermediate state (cf. Fig. 2). The idea was briefly mentioned   in  ref. \cite{blr}.
\begin{figure}[hbt]
	\vspace{-0.42cm}
	\hspace{-2cm}
	\resizebox{16cm}{!}{\includegraphics{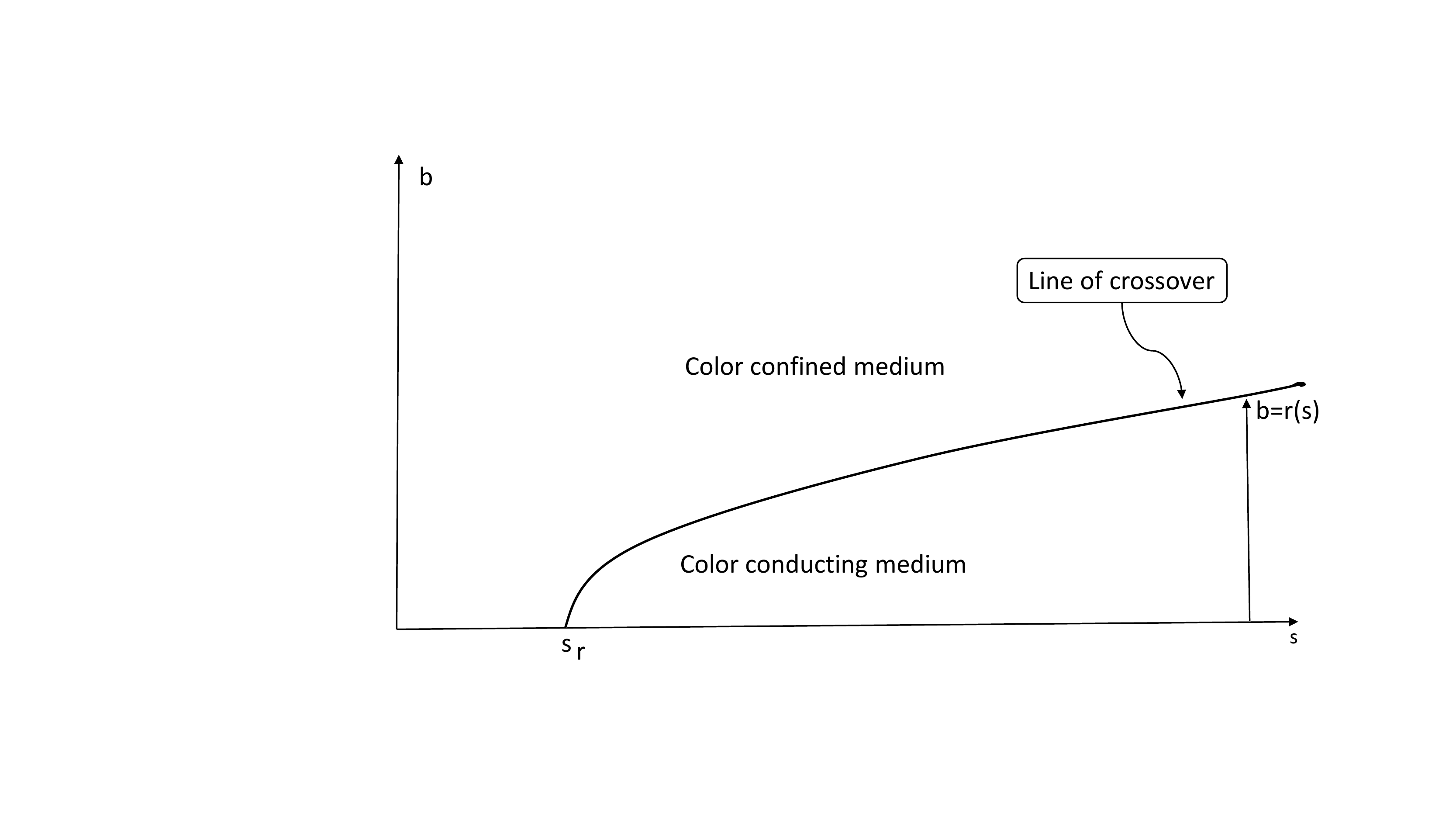}}		
	\vspace{-2cm}
	\caption{Two phases of hadronic matter corresponding the two scattering modes.}	
\end{figure}	 
Today, one can use analogy based  approach replacing an electromagnetic field of optics by a chromomagnetic field of QCD.  

 Formation of the color conducting medium in hadron collisions might also be responsible  for a number of collective effects in small systems such as correlations, anisotropic flows and others. Such effects can  arise  due to a ring-like shape  of the  impact--parameter  region responsible for the  multiparticle production processes.  Ring-like shape of this region is a result of the reflective scattering mode appearance and it implies  an important role of a coherent behavior of the deconfined matter formed under hadron interactions, resulting finally in the explicit collective effects  \cite{collect}.  Some of them have been mentioned above.

 We did not specify the nature of color conducting medium, namely, no need to discuss what kind of constituents form it -- massless colored current  quarks or massive colored constituent quarks -- thus, leaving aside a problem of chiral symmetry restoration scale versus the scale of color confinement. We  may expect  appearance of color conductivity in this medium as a result of deconfinement which occurs under  high-energy hadron collisions.
\section*{Acknowledgements}
We are grateful to V.A. Petrov for a stimulating discussion.

\end{document}